\documentclass[article, twocolumn]{elsarticle}
\biboptions{sort&compress}
\usepackage{ulem}
\usepackage{myterms}
\usepackage{lipsum}
\usepackage{amsmath}	
\usepackage{bm}		
\usepackage{amssymb}	
\usepackage{color} 		
\usepackage{graphicx} 	
\usepackage[T1]{fontenc}
\usepackage{times} 		
\usepackage{mathrsfs} 
\usepackage{hyperref} 
\usepackage{multicol}
\hypersetup{
    colorlinks=true,       
    linkcolor=blue,        
    citecolor=blue,        
    filecolor=magenta,     
    urlcolor=blue          
}
\usepackage[all]{hypcap} 
\usepackage[dvipsnames]{xcolor}
\usepackage[margin=1in]{geometry}

\newcommand{\Neff}{{N_\mathrm{eff}}}
\newcommand{\Smnu}{{\Sigma m_\nu}}
\renewcommand{\LCDM}{\Lambda\mathrm{CDM}}
\newcommand{\eLCDM}{\mathrm{e}\Lambda\mathrm{CDM}}

\begin{document}
\onecolumn 
\title{Unraveling the Dirac Neutrino with Cosmological and Terrestrial Detectors}
\author[1]{Peter Adshead}\ead{adshead@illinois.edu}

\author[2]{Yanou Cui}\ead{yanou.cui@ucr.edu}

\author[3]{Andrew J. Long}\ead{andrewjlong@rice.edu}

\author[2]{Michael Shamma\corref{cor1}}\ead{michael.shamma@email.ucr.edu}

\cortext[cor1]{Corresponding author}
\address[1]{Illinois Center for Advanced Studies of the Universe \& Department of Physics, 
University of Illinois at Urbana-Champaign, Urbana, IL 61801, USA.}
\address[2]{Department of Physics and Astronomy, University of California, Riverside, Riverside, CA 92521, USA.}
\address[3]{Department of Physics and Astronomy, Rice University, Houston TX 77005, USA.}

\date{\today}

\begin{abstract}

We point out a correlation between the effective number of relativistic degrees of species $\Neff$, the cosmologically measured $m_{\nu,\text{sterile}}^{\text{eff}}$, and the terrestrially measured neutrino mass sum and effective electron neutrino mass, $\Smnu$ and $m_{\nu_e}$, which arises in the Dirac neutrino hypothesis. If the neutrinos are Dirac particles, and if the active neutrinos' sterile partners were once thermalized in the early universe, then this new cosmological relic would simultaneously contribute to the effective number of relativistic species, $\Neff$, and also lead to a correlation between the cosmologically-measured effective sterile neutrino mass $m_{\nu,\text{sterile}}^\text{eff}$ and the terrestrially-measured active neutrino mass sum $\Smnu$. We emphasize that specifically correlated deviations in $\Neff \gtrsim 3$, $m_{\nu,\text{sterile}}^\text{eff}$ and $\Smnu$ above their standard predictions could be the harbinger revealing the Dirac nature of neutrinos. We provide several benchmark examples, including Dirac leptogenesis, that predict a thermal relic population of the sterile partners, and we discuss the relevant observational prospects with current and near-future experiments. If the correlation highlighted in this work is observed in future surveys, it could be interpreted as supporting evidence of Dirac neutrino masses.  

\end{abstract}



\maketitle

\newpageafter{keyword}
\clearpage

\twocolumn
\section{Introduction} 

The phenomenon of neutrino flavor oscillations~\cite{Pontecorvo:1957cp, Pontecorvo:1967fh, Maki:1962mu, Fukuda:1998mi, Ahmad:2001an} requires that at least two neutrinos are massive.  
However, the Standard Model (SM) predicts massless neutrinos~\cite{Weinberg:1967tq}, and therefore new physics is required to explain the origin of neutrino mass.  
Broadly speaking, neutrino mass models fall into two categories.  
The Majorana Neutrino Hypothesis (MNH) posits that neutrinos are their own antiparticles, and no new degrees of freedom are needed at the $O(m_\nu)$ scale~\cite{Weinberg:1979sa}.  
On the other hand, the Dirac Neutrino Hypothesis (DNH) introduces three new degrees of freedom that combine with the SM neutrinos to form three Dirac pairs of particles and antiparticles~\cite{Tanabashi:2018oca}. The DNH can be viewed as the limit where the Majorana mass terms in the MNH are zero. This limit is smooth, however: the quasi-Dirac (or ``pseudo-Dirac'') hypothesis (QDH) \cite{Wolfenstein:1981kw,Chang:1999pb,Balaji:2001fi,PhysRevD.80.073007} is that in which small deviations from vanishing Majorana mass terms give rise to light gauge-singlet sterile Majorana neutrinos which are near degenerate in mass with the left-handed (active) neutrinos. Determining whether neutrinos are Majorana or Dirac is of utmost importance for advancing our understanding of these elementary particles.  

A variety of experimental efforts are currently underway to test the Majorana or Dirac nature of the neutrinos.  
The MNH is particularly amenable to experimental probes, since these models generally predict lepton-number violation in two units, $\Delta L = \pm 2$, and thereby allow exotic nuclear reactions.  
For example, an observation of neutrinoless double beta decay~\cite{DellOro:2016tmg} would validate the MNH and imply that neutrinos are their own antiparticles~\cite{Schechter:1981bd}.

However, confirmation of the DNH is far harder to achieve.  
This is because the DNH's new degrees of freedom are light gauge-singlet sterile neutrinos, which do not interact with the $W^\pm$ and $Z$ bosons in the same way as the active neutrinos.  
Instead, their interactions are suppressed by the tiny Yukawa coupling $y_\nu = m_\nu / v_{ew} = O(10^{-12}) (m_\nu / 0.1 \eV)$, making their production and detection in the lab exceedingly unlikely.  
In this article we point out that a combination of terrestrial neutrino mass measurements and cosmological probes of neutrinos can yield evidence in favor of the DNH.  
If the neutrinos are Dirac particles, then for each ``active'' neutrino that we have measured, there must exist a \textit{precisely degenerate} ``sterile'' neutrino partner.  
In the minimal model, these sterile states are so weakly interacting that they are never produced in any significant abundance in the early universe~\cite{Shapiro:1980rr,Kolb:1981cx}, and they do not leave a detectable imprint on cosmological observables.  
However, there are many compelling beyond the SM scenarios in which the sterile states acquire a thermal distribution in the early universe and survive today as cosmological relics.  
We demonstrate how this leads to deviations in the cosmological neutrino observables (namely, $\Neff$ and $m_{\nu,\text{sterile}}^\text{eff}$) that are correlated with one another, as expected in general for eV-scale relics~\cite{DePorzio:2020wcz}, and also correlated with terrestrial neutrino observables (namely, $m_{\nu_e}$ and $\Smnu$), as also pointed out recently in \rref{Abazajian:2019oqj}.  
\qquad  \\ 

\section{Models with thermal sterile neutrinos}

In this section we discuss two examples in which Dirac neutrinos' sterile partners can reach a thermal or near-thermal abundance in the early universe: Dirac leptogenesis and theories with gauged $U(1)_{B-L}$ symmetry.  

We begin by reviewing why new physics is required to yield a substantial population of steriles~\cite{Shapiro:1980rr,Kolb:1981cx}.  
The simplest way to implement the Dirac neutrino hypothesis is via the Yukawa interaction
\begin{equation}\label{eq:minDiracL}
	\Lscr_{\nu}=Y_{\nu}^{ij}\bar{L}^{i}\tilde{H}\nu_{R}^j+\text{h.c.}
\end{equation} 
where $L = (\nu_L, e^-)$ is the left-handed lepton doublet, $H$ is the Higgs doublet, $\nu_R$ is the right-handed sterile neutrino, and $Y_{\nu}^{ij}$ is the matrix of Yukawa couplings.  
After electroweak symmetry breaking, $\nu_L$ and $\nu_R$ combine to give a Dirac fermion with mass $m_\nu\sim y_\nu v_{\text{ew}}$. Taking $m_\nu \sim 0.1 \eV$ requires $y_\nu \sim 10^{-12}$.  
Since this tiny Yukawa coupling is the sole interaction with the SM, the $\nu_R$ do not come into thermal equilibrium.  
Nevertheless, thermal freeze-in~\cite{Hall:2009bx} (see also \rref{Dodelson:1993je}) generates $\nu_R$ out of equilibrium via reactions such as 
$e^- \nu_L \to e^- \nu_R$
when the plasma temperature is $T \sim T_{ew} \sim 100 \GeV$.  
The predicted abundance is parametrically $\Omega_{\nu_R} \sim \rho_{\nu_R} / T_{ew}^4 \sim \langle \sigma v \rangle n_e n_{\nu_L} / H T_{ew}^3 \sim G_F^2 m_\nu^2 T_{ew} \Mpl$ where $\langle \sigma v \rangle \sim G_F^2 m_\nu^2$ is the thermally-averaged production cross section, $n_e \sim n_{\nu_L} \sim T_{ew}^3$ is the electron density, $H \sim T_{ew}^2 / \Mpl$ is the Hubble expansion rate, and $\Mpl \simeq 2.43 \times 10^{18} \GeV$ is the reduced Planck mass.  
Putting in numbers gives $\Omega_{\nu_R} \sim 10^{-8} \, (m_\nu / 0.1 \eV)^2$, which corresponds to an undetectably small population.  
 Thus if the neutrinos are Dirac and contribute detectably to the relativistic energy density in the early universe, physics beyond the SM is required. 

The relevant observable is the ``effective number of neutrino species" $\Neff$, defined as  the amount of relativistic energy density that is not in the form of cosmic microwave background (CMB) photons at decoupling measured in units of the energy carried by a SM neutrino. In the $\Lambda$-cold-dark-matter ($\Lambda$CDM) cosmology  $\Neff = \Neff^{(0)} \simeq 3.044$ \cite{Gnedin:1997vn, Mangano:2005cc, Grohs:2015tfy, Akita:2020szl, Escudero:2020dfa}. Provided that the sterile states were once in thermal equilibrium with the SM, the additional contribution to the radiation density is \cite{Brust:2013ova, Chacko:2015noa, Adshead:2016xxj, Abazajian:2019oqj} 
\begin{equation}
\Neff - \Neff^{(0)} \equiv \Delta \Neff\simeq(0.027 g_s)[106.75/g_\ast(T_\mathrm{dec})]^{\frac{4}{3}}
\end{equation}
where $g_s$ is the effective number of degrees of freedom (d.o.f.) of the new species, which is $2\times3\times 7/8=42/8$ in case of Dirac sterile neutrinos. $T_\mathrm{dec}$ is the temperature when the sterile states decouple, and $g_\ast(T_\mathrm{dec})$ is the total effective number of d.o.f's in the thermal bath just above $T_\mathrm{dec}$, which is 106.75 at weak scale temperature, $T \sim T_{ew}$, assuming SM field content only.  

New physics models that implement Dirac neutrinos generically introduce new fields and interactions that can efficiently produce and thermalize the sterile neutrinos \cite{Olive:1981ak,Chen:2015dka,Zhang:2015wua,Luo:2020sho,PhysRevD.102.035025}. We discuss two classes of models: \textit{Dirac leptogenesis} (LG) \cite{Dick:1999je,Murayama:2002je,Gu:2006dc,Gu:2012fg,Heeck:2013vha,Abel:2006nv,Ahn:2016hhq}, which  are motivated by explaining the observed baryon asymmetry using Dirac neutrinos, and  models with gauged Baryon minus Lepton number (B-L) symmetry \cite{Barger:2003zh,Anchordoqui:2011nh,Anchordoqui:2012qu,SolagurenBeascoa:2012cz,Nanda_2020}. A gauged $U(1)_{B-L}$ is particularly desirable in excluding $\Delta L=2$ Majorana mass terms, considering that their global symmetry counterpart may be undermined by quantum gravity \cite{Banks:2010zn,alvey2020axion}. The $U(1)_{B-L}$ may be broken by $\Delta L>2$ units for $m_Z'\neq0$.

In Dirac LG models, a new $SU(2)_L$ scalar doublet $\Phi$ decays out-of-equilibrium, and produce equal and opposite asymmetries of $\nu_R$ and $\nu_L$. These $\nu_R$ and $\nu_L$ are kept in thermal equilibrium in the early Universe through the scattering process mediated by $\Phi$. Nevertheless in order to avoid washing out the asymmetries produced in $\nu_R$ and $\nu_L$, these processes need to depart from equilibrium before, or around the time when Dirac LG is triggered by $\Phi$ decay. The exact time of $\Phi$ decay is model-dependent, but generally has to be before the electro-weak phase transition (EWPT) so that sphalerons can convert the lepton asymmetry to a baryon asymmetry. Therefore we find  that  $T_\mathrm{dec}$ should satisfy $T_\mathrm{dec}\gtrsim T_{\text{EWPT}}$ and $\Delta \Neff\sim 0.05-0.14$, depending on the number of new particles introduced in the model (e.g. whether in the framework of MSSM \cite{Murayama:2002je} or not \cite{Dick:1999je,Heeck:2013vha}). 

In gauged B-L models,  interactions mediated by $Z^\prime$ gauge bosons thermalize $\nu_R$ with the SM plasma via  s-channel processes such as $f\bar{f}\leftrightarrow\nu_R\bar{\nu}_R$. The decoupling of $\nu_R$ occurs below the $U(1)_{B-L}$ breaking scale. By comparing the interaction rate $\Gamma_{s}\sim g^{\prime4}T^5/m_{Z^\prime}^4$ and the Hubble expansion rate, we find the decoupling temperature $T_{\rm dec}\lesssim(m_{Z^\prime}/g^\prime M_\text{pl})^{4/3}M_\text{pl}$. 
Because these models extend the SM by up to three $\nu_R$ and $Z^\prime$ the total effective number of d.o.f can be as large as $g_\ast(T_\text{dec})\approx115$ and as low as $g_\ast(T_\text{dec})\approx75.25$, which translates to $\Delta N_\text{eff}\approx0.13-0.23$.

Additionally, a number of Dirac neutrino models are motivated by generating small neutrino masses through inclusion of additional Higgs doublets or mediator mass/loop suppression~\cite{Davidson:2009ha,Ma:2016mwh,Yao:2018ekp,Abazajian:2019oqj,PhysRevD.100.035041,han2020dirac}. Thus there are many reasons to expect abundantly populated and thermalized sterile neutrinos in the early universe. These sterile neutrinos would interact more weakly relative to their active counterparts and necessarily decouple earlier.

\qquad  \\ 
\section{Terrestrial and cosmological probes of neutrinos} 

Efforts are underway in the lab to measure the absolute neutrino mass scale.  
This can be parametrized by the effective electron neutrino mass, $m_{\nu_e} = [\sum_i m_i^2 |U_{ei}|^2]^{1/2}$, where $U_{ei}$ is the neutrino mixing matrix~\cite{Tanabashi:2018oca}.  
We are also interested in the sum of the three active neutrino masses, $\Smnu \equiv m_{\nu_1} + m_{\nu_2} + m_{\nu_3}$, which can be determined from the measured $m_{\nu_e}$ by knowing the squared mass splittings, the mixing angles, and mass ordering~\cite{Tanabashi:2018oca}. We choose to infer $\Smnu$ this way as the direct measurement of $m_{\nu_\mu}$ or $m_{\nu_\tau}$ has sensitivity of $O(0.1-10)$ MeV which is much worse than the related sensitivity from cosmological observations \cite{Assamagan:1995wb,PhysRevD.61.052002}. 
At present, the best experimental technique uses precision measurements of the tritium beta decay endpoint.  
Since a beta decay produces an (anti-)neutrino, which carries away an energy of at least $O(m_{\nu_e})$, the endpoint of the electron spectrum shifts downward for larger neutrino mass.  
Currently the best limits come from the KATRIN experiment~\cite{Osipowicz:2001sq}, which measures an effective neutrino mass squared value of $m_{\nu_e}^2 = (-1.0_{-1.1}^{+0.9}) \eV^2$, corresponding to an upper limit of $m_{\nu_e} < 1.1 \eV$ (90\% CL)~\cite{Aker:2019uuj}.  
As a constraint on the absolute neutrino mass scale, this is roughly $\Smnu \lesssim 3 \eV$. With more data, KATRIN is expected to constrain $m_{\nu_e} < 0.2 \eV$ (90\% C.L.) or measure the absolute neutrino mass ($5\sigma$) if it is larger than $m_{\nu_e} = 0.35 \eV$~\cite{Angrik:2005ep}, corresponding to $\Smnu = 1.05 \eV$ (same for either normal or inverted mass ordering).  
Concurrently the Project 8 experiment is under construction, and it aims to achieve a neutrino mass sensitivity at the level of $m_{\nu_e} = 0.04 \eV$ (90\% C.L.)~\cite{Esfahani:2017dmu}, which corresponds to $\Smnu = 0.14 \eV$ in the normally-ordered neutrino mass spectrum, and $0.099 \eV$ in the inverted one.   

At the same time, cosmological probes of relic neutrinos are becoming increasingly sensitive.  
During the epochs of Big Bang nucleosynthesis (BBN) and baryon acoustic oscillations (BAO), neutrinos were relativistic, and their effect on cosmology is primarily through $\Neff$. In particular, the presence of the sterile states  allows $\Neff > \Neff^{(0)}$.
 A larger $\Neff$ implies a larger radiation energy density, and therefore a larger Hubble parameter $H$.  During BBN, this changes the relationship between the expansion rate, $H$, and the temperature of the standard model plasma, and thus $\Neff$ at $T \sim 0.1 \MeV$ can be inferred from the primordial elemental abundances \cite{Yang:1978ge}.  At recombination, a larger expansion rate increases the angular scale of diffusion damping as compared with the acoustic scale, $\theta_d / \theta_s \sim H^{1/2}$~\cite{Hou:2011ec}, and so $\Neff$ at $T \sim 0.1$ eV can be inferred from measurements of the CMB power spectrum at small angular scales (high $\ell$).  
In particular, because neutrinos free-stream, they lead to a unique phase-shift of the CMB's acoustic peaks~\cite{Bashinsky:2003tk} that has now been detected~\cite{Follin:2015hya}.  
Currently, Planck restricts $\Delta \Neff < 0.3$ (95\% C.L.)~\cite{Aghanim:2018eyx}.  The constraint is slightly loosened if $\nu_R$ is interacting \cite{Chacko:2015noa, Baumann:2015rya, Brust:2017nmv} which is possible, \textit{e.g.}, in the gauged $U(1)_{B-L}$ scenario.
Future surveys, such as CMB Stage-IV, will constrain $\Delta \Neff \lesssim 0.06$ (95\% C.L.)~\cite{Abazajian:2016yjj, Abazajian:2019eic}.  
The CMB power spectra also rule out neutrinos that become non-relativistic around recombination, and require that each flavor of active neutrino satisfies $m_\nu < 1 \eV$~\cite{Aghanim:2018eyx}. 

At late times, the active neutrinos became nonrelativistic, and their energy density can be parametrized by $\Omega_\nu h^2 = \Sigma_i m_{\nu_i} n_{\nu_i} / (3 \Mpl^2 H_{100}^2)=\Smnu / 94 \eV$ where $H_{100} = 100 \km / \mathrm{sec} / \mathrm{Mpc}$, and $\Smnu\equiv\Sigma_i m_{\nu_i}$. In Planck's analysis, three species of neutrinos are assumed to be degenerate in mass, have zero-chemical potential, a Fermi-Dirac distribution with equal number densities $n_{\nu_i}$ and equal an temperature of $1.95~\text{K}$.
These cold neutrinos are free streaming on the scale $l_\mathrm{fs} \approx (250 \Mpc) (m_\nu / \mathrm{eV})^{-1}$~\cite{Dolgov:2008hz}.  
This suppresses the growth of structure, which leads to less lensing in the CMB and less structure today, probed by Lyman-$\alpha$ forest and BAO surveys.  
In particular, measurements of the gravitational lensing of the CMB as well as measurements of galaxy clustering are sensitive to the sum of the neutrino masses. 
This effect has not yet been observed, and future surveys will greatly increase the sensitivity~\cite{Abazajian:2016yjj}. Similarly, the contribution of additional,  sterile neutrinos can be constrained by Planck \cite{Aghanim:2018eyx}. Planck parameterizes the contribution of sterile neutrinos to the neutrino density as $m_{\nu,\text{sterile}}^\text{eff}/94~\text{eV}\equiv\Omega_{\nu,\text{sterile}}h^2$. The overall contribution to the neutrino energy density can be quantified as $\Omega_\nu^\text{tot}h^2=(1/94~\text{eV})(\Smnu+m_{\nu,\text{sterile}}^\text{eff})\equiv\Sigma m_\nu^\text{tot}/94~\text{eV}$.

A combination of CMB and large-scale structure (LSS) surveys put a strong upper bound on $\Smnu$~\cite{Vagnozzi:2017ovm}.  
The value of this bound is model dependent.  
For the 7-parameter model in which $\LCDM$ is extended by $\Smnu$ alone, a combination of Planck (temperature, polarization, lensing) and BAO gives  $\Smnu < 0.12 \eV$ (95\% C.L.)~\cite{Aghanim:2018eyx}. 
If the neutrino mass scale obeys the strong constraint, $\Smnu < 0.12 \eV$, then beta decay experiments like KATRIN will be unable to provide a laboratory measurement of the absolute neutrino mass scale.  
However, the cosmological limits are model dependent, and going beyond the restrictive framework of $\LCDM$ gives more freedom.  
For instance, the extended-$\LCDM$ model ($\eLCDM$)~\cite{DiValentino:2015ola} has up to $12$ free parameters, including $\Neff$, $\Smnu$, the dark energy equation of state, and the running of the scalar spectral index.  
In these models, the inferred upper limit is much weaker $\Smnu \lesssim 0.4 \eV$ (95\% C.L.), and remains consistent with laboratory limits from KATRIN.  
Additionally, Planck places model dependent bounds on $m_{\nu,\text{sterile}}^\text{eff}$. The model most relevant to this work is one with a minimal active neutrino mass hierarchy with $\Smnu=0.06~\text{eV}$ and a thermal sterile relic whose contribution can be matched to an effective neutrino mass $m_{\nu,\text{sterile}}^\text{eff}=(\Delta N_\text{eff}/3.044)^{3/4}m_\text{sterile}^\text{thermal}$ with $m_\text{sterile}^\text{thermal}$ the physical mass of the sterile neutrino. The combination of Planck and BAO gives the joint constraint of $m_{\nu,\text{sterile}}^\text{eff}<0.23 \eV$ and $\Neff<3.34$ (95\% C.L.) \cite{Aghanim:2018eyx}.
\qquad  \\ 
\section{Correlated terrestrial and cosmological observables} 

Let us now derive a correlation between the cosmological neutrino observables, $\Neff$, $\Smnu^\text{tot}$, and $m_{\nu,\text{sterile}}^\text{eff}$, and the terrestrial neutrinos observable, $\Smnu$.  
The result Eq. \pref{eq:master_formula} appears in \rref{Abazajian:2019oqj} without derivation.  
We present the derivation here, both for readers who are unfamiliar with cosmological relic calculations and to emphasize the special role played by the degeneracy of active and sterile neutrinos in the DNH.  
We show that this correlation arises if the neutrinos are Dirac and if there is a thermal population of the sterile states.  
See also \rref{DePorzio:2020wcz} for a discussion of general eV-scale relics and their impact on cosmological observables, and see Refs.~\cite{Olive:1981ak,Abazajian:2019oqj} for a discussion of Dirac neutrinos in cosmology.  

First we consider the relic background of active neutrinos.  
The active neutrinos were initially in thermal equilibrium with the primordial plasma, and they decoupled at a time $t_{a,\mathrm{dec}}$ when the scale factor was $a_{a,\mathrm{dec}} = a(t_{a,\mathrm{dec}})$ and the plasma temperature was $T_{a,\mathrm{dec}} = T(t_{a,\mathrm{dec}}) \simeq \mathrm{MeV}$.  
Since the neutrino masses are $\lesssim O(\mathrm{eV})$, the active neutrinos were relativistic at the time of decoupling.  
After decoupling, the active neutrinos cooled due to the cosmological expansion and their temperature decreased as $T_{a,\mathrm{dec}} \, [ a(t) / a_{a,\mathrm{dec}} ]^{-1}$.  
Eventually the active neutrinos became non-relativistic at a time $t_{a,\mathrm{nr}}$ when $a_{a,\mathrm{nr}} = a(t_{a,\mathrm{nr}})$ such that their mass can be related to the decoupling temperature by $m_a = T_{a,\mathrm{dec}} \ ( a_{a,\mathrm{nr}} / a_{a,\mathrm{dec}} )^{-1} $.  
The formulas here hold approximately as long as the mass splitting is smaller than the total mass, so that the three active neutrinos become non-relativistic within $O(1)$ Hubble times of one another.  
We can write the energy density in the active neutrinos as~\cite{Kolb:1990} 
\begin{align}\label{eqn:activedensity}
	\rho_{a} \approx \begin{cases}
	6 \frac{7}{8} \frac{\pi^2}{30} T_{a,\mathrm{dec}}^4 \! \left( \frac{a(t)}{a_{a,\mathrm{dec}}} \right)^{\! -4} 
	& \! \! \! \! \! , \ t_{a,\mathrm{dec}} < t < t_{a,\mathrm{nr}} \\ 
	2 \frac{7}{8} \frac{\pi^2}{30} T_{a,\mathrm{dec}}^3 \! \left( \frac{a(t)}{a_{a,\mathrm{dec}}} \right)^{\! -3}\Smnu
	& \! \! \! \! \! , \ t_{a,\mathrm{nr}} < t 
	\end{cases} 
\end{align}
where the assumptions of instantaneous decoupling and non-relativistic transition introduce a $\lesssim O(1)$ uncertainty. 

Next we turn our attention to the relic background of sterile neutrinos.  
We assume that the sterile neutrinos were once in thermal equilibrium with the plasma, and that they decoupled at a time $t_{s,\mathrm{dec}}$ such that $a_{s,\mathrm{dec}} = a(t_{s,\mathrm{dec}})$ and $T_{s,\mathrm{dec}} = T(t_{s,\mathrm{dec}})$.  
We are interested in $t_{s,\mathrm{dec}} < t_{a,\mathrm{dec}}$ and $T_{s,\mathrm{dec}} > T_{a,\mathrm{dec}} \simeq \mathrm{MeV}$ so that the sterile neutrinos decoupled before the active neutrinos.  
In that case, they will not share the entropy injections from the decoupling of other Standard Model species, and we expect the sterile neutrinos to be colder than the active neutrinos at any given time.  
This means that the sterile neutrinos became nonrelativistic before the active neutrinos.  
Let $t_{s,\mathrm{nr}}$ be the time when the sterile neutrinos became non-relativistic with $a_{s,\mathrm{nr}} = a(t_{s,\mathrm{nr}})$ such that $m_s = T_{s,\mathrm{dec}} \ ( a_{s,\mathrm{nr}} / a_{s,\mathrm{dec}} )^{-1}$.  
Here we write the sum of sterile neutrino masses as $\Sigma_i m_{s,i}$, which must equal the sum of active neutrino masses $\Smnu$ in the DNH, but the formulas written here also apply for a more general eV-scale relic with $\Sigma m_{s,i} \neq \Smnu$.
Then the energy density of the sterile neutrinos is written as 
\begin{align}\label{eqn:steriledensity}
	\rho_{s} \approx \begin{cases}
	6 \frac{7}{8} \frac{\pi^2}{30} T_{s,\mathrm{dec}}^4 \! \left( \frac{a(t)}{a_{s,\mathrm{dec}}} \right)^{\! -4} 
	& \! \! \! \! \! , \ t_{s,\mathrm{dec}} < t < t_{s,\mathrm{nr}} \\ 
	2 \frac{7}{8} \frac{\pi^2}{30} T_{s,\mathrm{dec}}^3 \! \left( \frac{a(t)}{a_{s,\mathrm{dec}}} \right)^{\! -3}\Sigma_{i} m_{s,i} 
	& \! \! \! \! \! , \ t_{s,\mathrm{nr}} < t 
	\end{cases}.
\end{align}
Note that $\rho_{s} \leq \rho_{a}$ since $T_{s,\mathrm{dec}} a_{s,\mathrm{dec}} \leq T_{a,\mathrm{dec}} a_{a,\mathrm{dec}}$.  

The cosmological observables are $\Neff$ and $\Sigma m_\nu^\text{tot}$ where $\Sigma m_\nu^\text{tot}$ includes contributions from the active neutrinos and eV-scale relics. We can write 
\begin{equation}\label{eq:Neff_Smnu}
\begin{split}
	\Neff & = \Kcal_N \bigl( \rho_{a} + \rho_{s} \bigr) \bigr|_{t = t_\text{\sc cmb}} 
	 \quad \text{and} \quad \\
	\Sigma m_\nu^\text{tot} & = \Kcal_m \bigl( \rho_{a} + \rho_{s} \bigr) \bigr|_{t = t_0} ,
\end{split} 
\end{equation}
where the numerical coefficients, $\Kcal_N$ and $\Kcal_m$, are independent of the mass hypothesis.  
It is also useful to define $\Neff^{(0)}$ and $\Smnu$ by setting $\rho_s = 0$ in \eref{eq:Neff_Smnu}.  
We know that $\Neff^{(0)} \simeq 3.044$ counts the three active neutrino flavors~\cite{Akita:2020szl,Escudero:2020dfa}, and $\Smnu$ is the sum of the three active neutrino masses.  

Now let us consider the ratios
\begin{equation}
\begin{split}
	\frac{\Neff}{\Neff^{(0)}} & = 1 + \frac{\rho_{s}}{\rho_{a}} \Bigr|_{t = t_\text{\sc cmb}} 
	\quad \text{and} \quad \quad \\
	\frac{\Sigma m_\nu^\text{tot}}{\Smnu} & = 1 + \frac{\rho_{s}}{\rho_{a}} \Bigr|_{t = t_0} 
	\com
\end{split}
\end{equation}
whose deviations from unity parametrize the effect of the sterile neutrinos.  
We evaluate these expressions using the formulas for $\rho_{a}$ and $\rho_{s}$ that appear above.  
We know that the active neutrinos are relativistic at $t_\text{\sc cmb}$ and that they are nonrelativistic at $t_0$.  
Since the sterile neutrinos are colder than the active ones, we know that they are also nonrelativistic at $t_0$.  
We \textit{assume} that the sterile neutrinos are relativistic at $t_\text{\sc cmb}$, and this assumption can be checked in a particular model for sterile neutrino production.  
Using Eqns.\ \eqref{eqn:activedensity} and \eqref{eqn:steriledensity}, we have 
\begin{equation}\label{eq:master0}
\begin{split}
	\frac{\Neff}{\Neff^{(0)}} & = 1 + \left( \frac{a_{s,\mathrm{dec}} T_{s,\mathrm{dec}}}{a_{a,\mathrm{dec}} T_{a,\mathrm{dec}}} \right)^4 
	\quad \text{and} \quad \\ 
	\frac{\Sigma m_\nu^\text{tot}}{\Smnu} & = 1 + \frac{\Sigma_i m_{s,i}}{\Smnu} \left( \frac{a_{s,\mathrm{dec}} T_{s,\mathrm{dec}}}{a_{a,\mathrm{dec}} T_{a,\mathrm{dec}}} \right)^3  
	\per
\end{split}
\end{equation}
For a general eV-scale relic, the mass ratio $\Sigma_i m_{s,i}/\Smnu$ may be different from unity, but if the eV-scale relics are the sterile (Dirac) partners to the active neutrinos, then we must have $\Sigma_i m_{s,i} / \Smnu = 1$, and we find
\begin{equation}\label{eq:master_formula}
	\left( \frac{\Neff}{3.044} - 1 \right)^{1/4} 
	=\left( \frac{\Sigma m_\nu^\text{tot}}{\Smnu} - 1 \right)^{1/3}=\left(\frac{m_{\nu,\text{sterile}}^\text{eff}}{\Smnu}\right)^{1/3} 
	\per 
\end{equation}
This is the correlation that we set out to derive. 
An equivalent expression appears in \rref{Abazajian:2019oqj}. If we assume a particular mass hierarchy, the active neutrino mass sum can be used to rewrite \eref{eq:master_formula} in terms of $m_{\nu_e}$. 

To parameterize this expression, recall that the comoving entropy density at time $t$ is $a(t)^3 \, s(t) = (2\pi^2/45) \, g_{\ast S}(t) \, a(t)^3 \, T(t)^3$, where $g_{\ast S}(t)$ is the effective number of relativistic species~\cite{Kolb:1990}.  
If the comoving entropy density is conserved between times $t_{s,\mathrm{dec}}$ and $t_{a,\mathrm{dec}}$, then we can write 
\begin{align}\label{eqn:entropyexpansionrel}
	\frac{a_{s,\mathrm{dec}} T_{s,\mathrm{dec}}}{a_{a,\mathrm{dec}} T_{a,\mathrm{dec}}} = \left( \frac{g_{\ast S}(t_{a,\mathrm{dec}})}{g_{\ast S}(t_{s,\mathrm{dec}})} \right)^{1/3} 
	\com
\end{align}
which is the ratio that appears in \eref{eq:master0}.  

\begin{figure}[t]
\begin{center}
\includegraphics[width=0.45\textwidth]{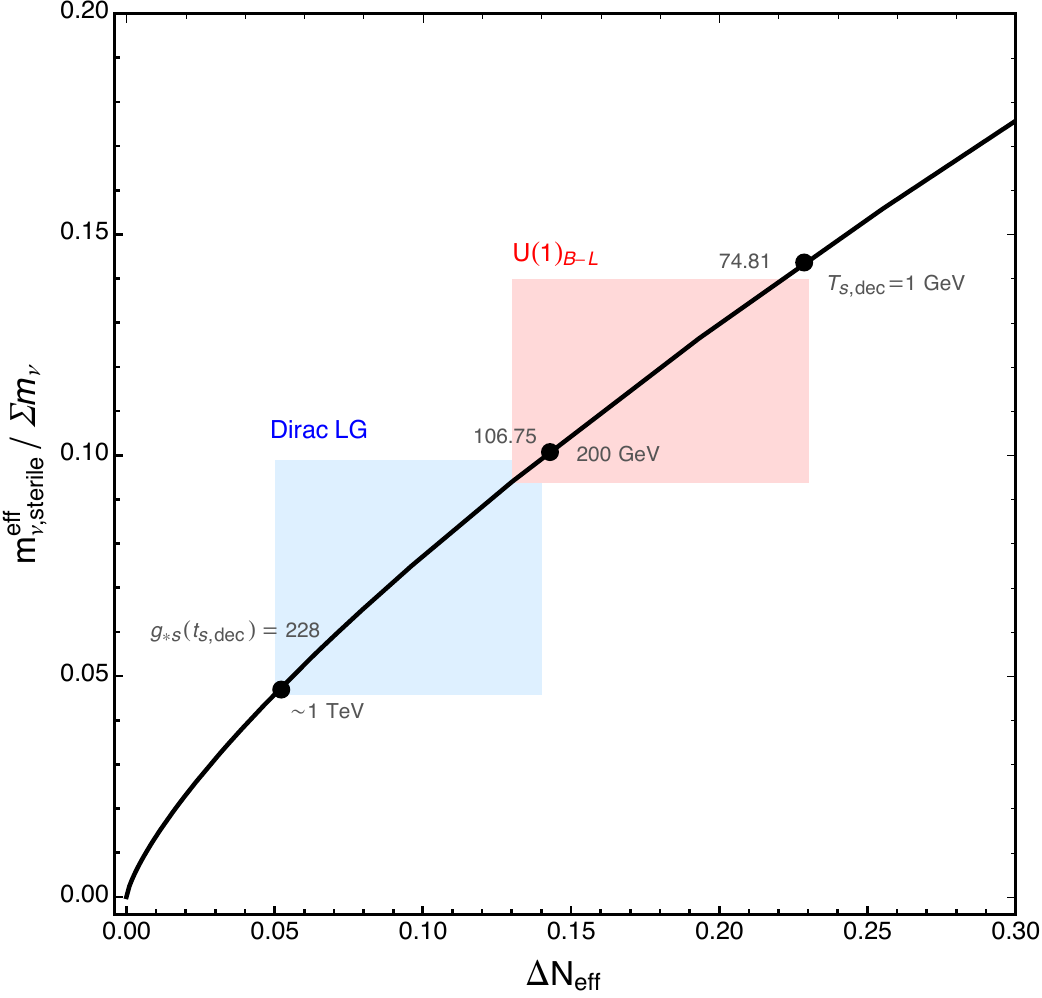} 
\caption{\label{fig:Neff_Smnu}
The effective number of neutrino species $\Neff$ and the effective sum of neutrino masses $m_{\nu,\text{sterile}}^\text{eff}$ are related through \eref{eq:master_formula}.  Each point along the curve corresponds to a different decoupling temperature for the sterile neutrinos $T_{s,\mathrm{dec}}$ (marked on the left side of the curve), which is also parametrized by $g_{\ast S}$ at that time~\cite{Husdal:2016haj} (marked on the right side of the curve). The shaded regions show the ranges for $m_{\nu,\text{sterile}}^\text{eff}$ as predicted by Eq. 8 in our work using the ranges for $\Delta N_\text{eff}$ in Dirac Leptogenesis (blue) and $U(1)_{B-L}$ (red) as given in Sec. 2.
}
\end{center}
\end{figure}

\qquad  \\ 
\section{Results and discussion}  

The relation in \eref{eq:master_formula} implies a correlation between the cosmological neutrino observables, $\Neff$ and $m_{\nu,\text{sterile}}^\text{eff}$, and $\Smnu$ as inferred from terrestrial observable $m_{\nu_e}$.  
We plot this relation in \fref{fig:Neff_Smnu}, along with various benchmark points derived using \eref{eqn:entropyexpansionrel}. The colored boxes in Fig. \fref{fig:Neff_Smnu} shows the range of predicted for $m_{\nu,\text{sterile}}^\text{eff}/\Smnu$ assuming the values of $0.05<\Delta\Neff<0.14$ in Dirac Leptogenesis (blue) and $0.13<\Delta\Neff<0.23$ in $U(1)_{B-L}$ (red) models.
Recall that $g_{\ast S}(t_{a,\mathrm{dec}}) \approx 10.75$ is the effective number of relativistic species when the active neutrinos decouple near $T_{a,\mathrm{dec}} \simeq 1$ MeV, and it is $g_{\ast S}(t_{s,\mathrm{dec}}) > 10.75$ when the sterile neutrinos decouple.  
If $g_{\ast S}(t_{s,\mathrm{dec}}) \gg g_{\ast S}(t_{a,\mathrm{dec}})$ then the sterile neutrinos are much colder than the active ones at any given time, implying $\Neff \approx 3.044$ and $\Sigma m_\nu^\text{tot} \approx \Smnu$.  
Alternatively if $g_{\ast S}(t_{s,\mathrm{dec}}) \approx g_{\ast S}(t_{a,\mathrm{dec}})$ then the sterile neutrinos decouple at almost the same time as the active ones, implying $\Neff \approx 6$ and $\Sigma m_\nu^\text{tot} \approx 2 \Smnu$.  As detailed in Sec. 2, BSM scenarios posit new interactions for eV-scale sterile neutrinos which would allow them to decouple later and contribute to $\Neff$. In particular, in the most optimistic models of Dirac LG and gauged $B-L$ which predict additional contributions to $\Neff$ as large as $\Delta N_{\rm eff}\approx0.14$. According to the correlation of \eref{eq:master_formula}, these models would also contribute to $\Sigma m_\nu^\text{tot}$ that would lie on the curve in \fref{fig:Neff_Smnu} with $\Sigma m_\nu^\text{tot}\approx1.15$. 

\begin{figure}[t!]
\begin{center}
\includegraphics[width=0.45\textwidth]{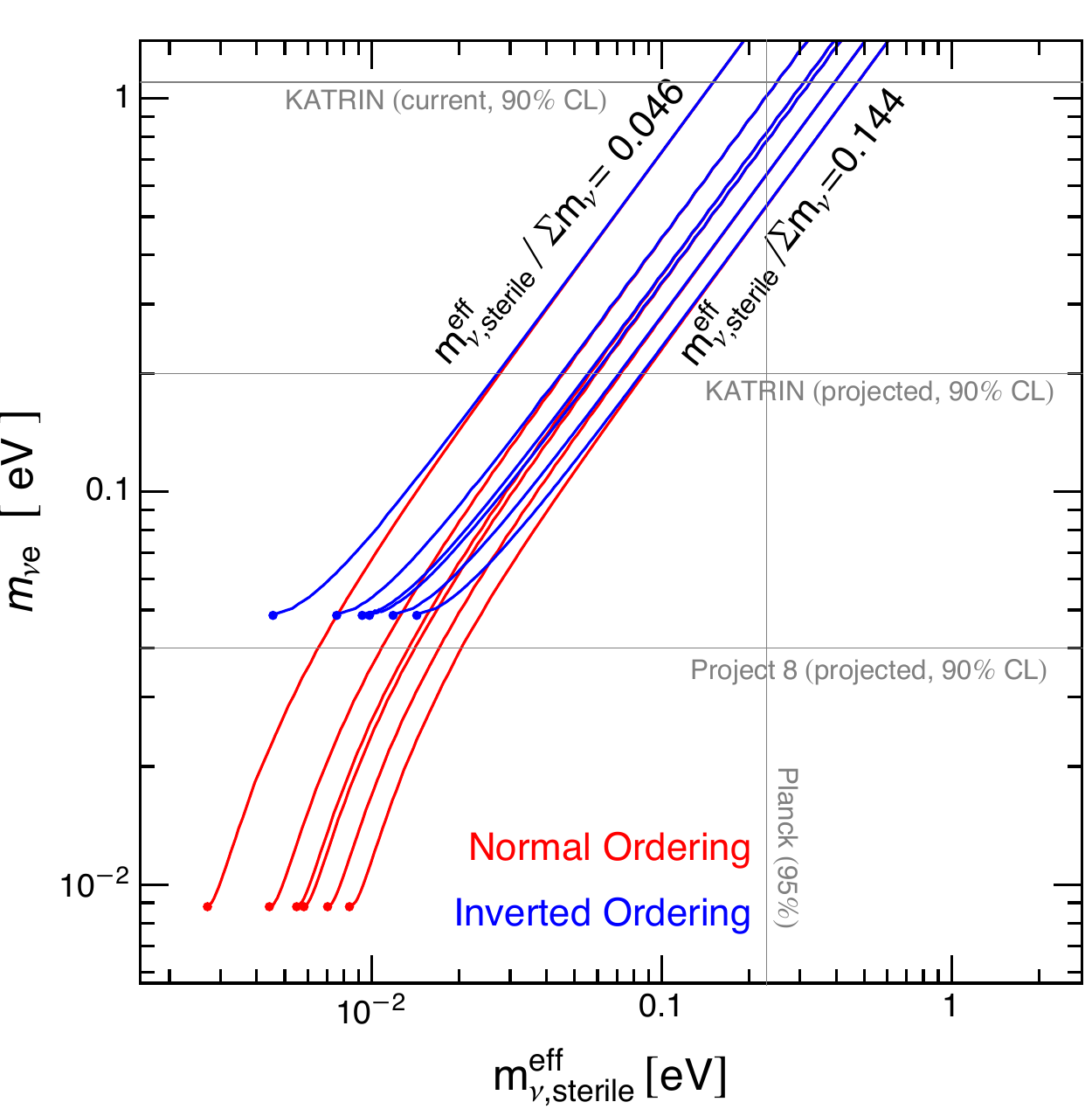} 
\caption{\label{fig:Smn_mne}
The effective sterile neutrino mass $m_{\nu,\text{sterile}}^\text{eff}$ and the effective electron-type neutrino mass $m_{\nu_e}$ are shown here with overlaid curves corresponding to different values of $m_{\nu,\text{sterile}}^\text{eff}/ \Smnu = 0.046, 0.076, 0.094, 0.099, 0.120$ and $0.144$.  These values correspond with $\Delta\Neff=0.05, 0.098, 0.13, 0.14, 0.18$ and $0.23$, respectively. Note that $0.05<\Delta\Neff<0.23$ is the range predicted by both benchmark models discussed in Section 2. KATRIN and Project 8 sensitivities are taken from Refs.~\cite{Osipowicz:2001sq,Angrik:2005ep,Esfahani:2017dmu,Aker:2019uuj} while the bound on $m_{\nu,\text{sterile}}^\text{eff}$ is taken from \cite{Aghanim:2018eyx}.  The red and blue curves correspond to the normal and inverted mass ordering, respectively, and the regime where they merge (upper-right corner) corresponds to the quasi-degenerate regime. }  
\end{center}
\end{figure}

In \fref{fig:Smn_mne} we show a parameter space in which the effective sterile neutrino mass, $m_{\nu,\text{sterile}}^\text{eff}$, and the effective electron-type neutrino mass, $m_{\nu_e}$, are varied.  
We overlay curves of constant $m_{\nu,\text{sterile}}^\text{eff}/ \Smnu$ from $0.046$ to $0.144$, which correspond to constant $\Neff$ from $3.094$ to $3.188$ through \eref{eq:master_formula} and sterile neutrino decoupling time. These curves of constant $m_{\nu,\text{sterile}}^\text{eff}/ \Smnu$ also represent the predictions of the DNH with a once-thermalized population of sterile partners in the benchmark models of Sec. 2.
The mapping from $m_{\nu_e}$ to $m_{\nu,\text{sterile}}^\text{eff}$ depends on the neutrino mass hierarchy, and we show both normal ordering (red) and inverted ordering (blue). The horizontal lines correspond to current limits and projected sensitivities from KATRIN and Project 8 on the active neutrino mass sum. The vertical line corresponds to the Planck limit on $m_{\nu,\text{sterile}}^\text{eff}<0.23 \eV$ \cite{Aghanim:2018eyx}. Since we are interested in probing the Dirac neutrino case, the latter limit on $m_{\nu,\text{sterile}}^\text{eff}$ is meant to indicate the typical sensitivity, but it does not apply directly to our model for which the active neutrinos and their sterile partners have the same mass, so their masses must vary together in a specific mass hierarchy. However, through the curves parameterized by $\Delta N_\text{eff}$, the limit on $m_{\nu,\text{sterile}}^\text{eff}$ can be used to constrain $m_{\nu_e}$ in the DNH via the correlation of \eref{eq:master_formula}. For example for the curve labeled $m_{\nu,\text{sterile}}^\text{eff}/\Smnu=0.144$, corresponding to $\Delta N_\text{eff}=0.23$, the correlation of \eref{eq:master_formula} requires $m_{\nu_e}\lesssim0.54 \eV$. Although this is above the projected sensitivity of KATRIN, the bound extrapolated from \eref{eq:master_formula} is sufficiently stronger than the current bound.
The curves terminate at a point (indicated by a dot) where the lightest neutrino mass is vanishing. If the neutrino masses are very hierarchical, the connection with $\Neff$ in \eref{eq:master_formula} holds as long as the three flavors of sterile neutrino decouple at the same temperature as one another, but at higher temperature than that of the three flavors of active neutrinos.

Future cosmological surveys will have much greater sensitivity to $\Neff$, $\Smnu$, and $m_{\nu,\text{sterile}}^\text{eff}$, while next-generation neutrino experiments aim to deliver a precision measurement of $m_{\nu_e}$ and $\Smnu$.  
How will this data inform our understanding of neutrinos and their role in cosmology?  
If these experiments discover that $\Neff \gtrsim 3.044$ and $\Sigma m_\nu^\text{tot} \gtrsim \Smnu$,
this would provide strong evidence for the presence of a new cosmological relic with mass $m = O(0.1) \eV$~\cite{DePorzio:2020wcz}.  
What are the candidates for this eV-scale relic and how can we distinguish them with laboratory tests?  
The coincidence of $m$ and $m_\nu = O(0.1) \eV$ suggests a specific connection between the new relic and the known active neutrinos.  

In this work, we have argued that the Dirac neutrino hypothesis provides a compelling explanation for eV-scale relics if there is a thermal population of sterile neutrinos.  
This hypothesis together with the assumption that sterile Dirac partners are thermalized predicts the correlation in Eq. \pref{eq:master_formula} between $\Neff$ and the masses. If this correlation were measured with future surveys, it would be evidence for the Dirac neutrino hypothesis as it requires $\Neff$, $m_{\nu,\text{sterile}}^\text{eff}$, and $\Smnu$ to obey \eref{eq:master_formula}. Although it is possible for other new physics to mimic the specific signal of Eq. \pref{eq:master_formula}, this is unlikely as this signal is reliant specifically on the eV-relic being mass degenerate with the active neutrino of the SM. Thus, if future cosmological and terrestrial surveys such as CMB-S4 and Project-8 measured this correlation and $\Delta\Neff\sim0.05~(0.13)$ and $\Smnu\sim0.046~(0.094)$, this would point to the veracity to models of Dirac LG ($U(1)_{B-L}$) and would motivate dedicated analyses.
These measurements would provide strong evidence for a thermal population of relic sterile neutrinos. 
Additionally, measurements of $\Neff$ and $\Sigma m_\nu^\text{tot}/ \Smnu$ could be used to infer the sterile decoupling temperature through \eref{eq:master0}, which would provide additional information, helping to understand how the steriles were populated. For example, 
Moreover, these sterile neutrinos would survive in the universe today.  
Laboratory efforts to directly detect the relic neutrino background~\cite{Betts:2013uya,Baracchini:2018wwj} could also uncover the presence of the sterile neutrinos~\cite{Long:2014zva,Chen:2015dka,Zhang:2015wua,Betti:2019ouf}, providing an additional handle on this scenario.  
When the terrestrial and cosmological observables are taken together, they could be used to support the Dirac neutrino hypothesis.

Utilizing a correlation between the cosmological variables $\Neff$, $m_{\nu,\text{sterile}}^\text{eff}$ and the terrestrially measured variables $\Smnu$ , $m_{\nu_e}$ , we have proposed a diagnostic test for the Dirac Neutrino Hypothesis.  If future data returns a positive result for this test, it will be necessary to follow up with a dedicated analysis.  In particular, the DNH is distinguished from generic eV-scale sterile neutrinos since the mass of the sterile partners must equal the mass of the active neutrinos.  This model is not fully captured by existing studies for which $m_{\nu, \text{sterile}}^\text{eff}$ is varied while $\Smnu$ is held fixed or for which $\Neff$ and $\Smnu$ are varied with $m_{\nu,\text{sterile}}^\text{eff}$ held fixed. In this way a combination of the diagnostic test and follow-up studies may provide strong evidence for the Dirac neutrino hypothesis.


\section*{Acknowledgments}
\textit{
The authors thank Kev Abazajian and Julian Heeck for comments on the manuscript. The work of PA is supported in part by the US Department of Energy Grant No.\ DE-SC0015655. YC is supported in part by the US Department of Energy grant DE-SC0008541. PA, YC and AL thank the hospitality of the Kavli Institute for Theoretical Physics while the work was in progress, which is supported in part by the National Science Foundation under Grant No.\ NSF-PHY-1748958.  }

\bibliography{draft}

\end{document}